\documentclass[journal]{IEEEtran}

\usepackage{cite}
\usepackage{graphicx}
\usepackage{amsmath}
\usepackage{amssymb}
\usepackage{amsfonts}
\usepackage{mathtools}
\usepackage[ruled,vlined,linesnumbered]{algorithm2e}
\usepackage{algorithmic}
\usepackage{url}
\usepackage{float}

\usepackage[table,svgnames,hyperref,HTML]{xcolor}
\definecolor{green}{HTML}{1A8033}
\definecolor{blue}{HTML}{4DB3E6}
\definecolor{red}{HTML}{E64D4D}

\usepackage{tabularx}
\usepackage{caption}
\usepackage{multirow}
\usepackage{makecell}
\usepackage{booktabs}

\usepackage{setspace}
\begin{document}

\title{WiFo-E: A Scalable Wireless Foundation Model for End-to-End  FDD Precoding in Communication Networks}

\author{
Weibo Wen,~\IEEEmembership{Graduate Student Member,~IEEE}, Shijian Gao,~\IEEEmembership{Member,~IEEE}, Haotian Zhang,~\IEEEmembership{Graduate Student Member,~IEEE}, Xiang Cheng,~\IEEEmembership{Fellow,~IEEE} and Liuqing Yang,~\IEEEmembership{Fellow,~IEEE} 
  \thanks{Weibo Wen, Haotian Zhang and Xiang Cheng are with the State Key Laboratory of Photonics and Communications, Peking University, Beijing 100871, P. R. China (email: weber@stu.pku.edu.cn; haotianzhang@stu.pku.edu.cn; xiangcheng@pku.edu.cn).}
  \thanks{Shijian Gao is with the Internet of Things Thrust, The Hong Kong University of Science and Technology (Guangzhou), Guangzhou 511400, P. R. China (e-mail: shijiangao@hkust-gz.edu.cn).}
  \thanks{Liuqing Yang is with the Internet of Things Thrust and Intelligent Transportation Thrust, The Hong Kong University of Science and Technology (Guangzhou), Guangzhou 510000, P. R. China (email: lqyang@ust.hk).}
}

\markboth{Journal of \LaTeX\ Class Files,~Vol.~X, No.~X, January~2026}{Journal of \LaTeX\ Class Files,~Vol.~X, No.~X, January~2026}%

\maketitle

\begin{abstract}

Accurate precoding in massive multiple-input multiple-output (MIMO) frequency-division duplexing (FDD) systems relies on efficient channel state information (CSI) acquisition. End-to-end learning frameworks improve performance by jointly optimizing this process, but they lack scalability and fail to generalize across different system configurations, such as varying numbers of antennas and users. To overcome this limitation, we introduce WiFo-E, a wireless foundation model designed for scalable end-to-end precoding. WiFo-E employs multi-task pretraining on a diverse set of configurations to learn transferable representations of underlying wireless principles. Central to the model is a sparse Mixture-of-Experts (MoE) Transformer architecture, which mitigates task interference and enhances training efficiency by activating specialized parameter subsets adaptively. Extensive simulations demonstrate that WiFo-E outperforms conventional per-configuration training and shows strong generalization to unseen system configurations, providing a flexible and efficient foundation for adaptive massive MIMO precoding.

\end{abstract}


\bstctlcite{BSTcontrol}

\section{Introduction}

\IEEEPARstart{M}{assive} multiple-input multiple-output (MIMO) has significantly enhanced the performance of fifth-generation (5G) wireless networks, thereby providing gains in spectral efficiency and system capacity~\cite{Cheng_Zhang_Zhang_Gao_Li_Huang_Bai_Yang_Zheng_Yang_2024}. As sixth-generation (6G) communication systems demand even higher data rates, lower latency, and higher connection density, MIMO technologies are expected to play an even more critical role in meeting these requirements. However, to exploit the full potential of MIMO, it is critical to acquire accurate channel state information (CSI) with low overhead and to compute the optimal precoder for transmission~\cite{luo2018channel}\cite{wang2024deep}. Typical massive multiple-input multiple-output (MIMO) systems operate in time-division duplexing (TDD) mode, where the base station (BS) receives uplink pilot signals and exploits channel reciprocity to acquire CSI for downlink precoding. In addition, many existing wireless networks operate in frequency-division duplexing (FDD) mode, where uplink--downlink reciprocity does not hold. In FDD systems, the downlink channel must be estimated and then fed back to the BS for precoding. Designing an end-to-end pipeline to achieve optimal precoding is therefore essential for the widespread adoption of FDD massive MIMO in future communication networks.

In recent years, deep learning (DL) has emerged as a promising alternative to traditional rule-based approaches in wireless communication systems~\cite{huang2019deep, hu2021distributed, Wang_Li_He_Jiang_Ruby_Ji_Leung_2022, Alkhateeb_Leus_Heath_2015}. Deep neural networks (DNNs) have been applied to physical-layer tasks such as channel estimation~\cite{luo2018channel}, precoding~\cite{wang2024deep}, and signal detection in large-scale MIMO systems. Early studies typically designed and trained modules such as channel estimation, CSI feedback, and precoding in isolation, which were subsequently integrated into communication systems as separate components. Although such modular designs perform well on individual tasks, their training objectives are inconsistent. For instance, channel estimation and feedback aim to minimize the normalized mean squared error (NMSE), while precoding is usually designed to maximize the sum rate. When the ultimate objective is downlink precoding in an FDD massive MIMO system, estimation and feedback modules introduce considerable redundancy, since precoding is more concerned with users' spatial directions and correlations than with reconstructing every channel matrix entry.  This misalignment results in suboptimal end-to-end performance. Moreover, training modules in isolation leads to error propagation across the system, whereby imperfections in early-stage outputs, such as estimation or feedback, are amplified in downstream modules, ultimately degrading the overall performance of the pipeline. To address these issues, the authors in~\cite{Sohrabi_Attiah_Yu_2021} proposed an end-to-end FDD multi-user MIMO precoding framework that jointly optimizes multiple modules, including downlink pilot design, channel feedback, and precoding. Under the same feedback overhead, this end-to-end framework significantly outperforms conventional designs based on independent training. Subsequent studies have explored several extensions of this end-to-end precoding framework. For instance, the authors in~\cite{Jang_Lee_Kim_Lee_2022} proposed an adaptive codebook-based feedback scheme to reduce channel state reporting overhead. In~\cite{Hu_Cai_Kang_Yu_Hoydis_Eldar_2022}, the architecture was extended from fully digital precoding to hybrid precoding. Furthermore, the authors in~\cite{Kang_Hu_Cai_Yu_Hoydis_Eldar_2022} incorporated temporal correlation, enabling the framework to handle time-varying channels across multiple transmission slots.

Despite the performance improvements achieved by end-to-end learning, the mappings learned by data-driven models are tightly coupled to the specific system configurations (e.g., array size and number of users) seen during training. This creates a scalability challenge: a precoding model trained for one configuration often fails to generalize to others, as practical transceivers operate under diverse, time-varying settings~\cite{ma2022learn, bai2020precoding}. Several solutions have been proposed to address this challenge. In~\cite{Sohrabi_Attiah_Yu_2021}, dedicated DNNs are trained and stored for each possible number of users. However, this approach incurs substantial training and storage overhead as configurations proliferate. To improve scalability across configurations, subsequent work~\cite{Hu_Cai_Kang_Yu_Hoydis_Eldar_2022} fixed the dimensionality of the CSI and padded missing entries to allow joint training under multiple configurations. Moreover, models with inherently flexible architectures, such as graph neural networks (GNNs)~\cite{Zhao_Wu_Ma_Yang_2024, zhu2025scalable} and Transformers~\cite{dehghani2023patchnpacknavit,Vaswani_Shazeer_Parmar_Uszkoreit_Jones_Gomez_Kaiser_Polosukhin_2017}, have been adopted to accommodate variable input dimensions via graph-based modeling or attention mechanisms, for example, in multi-user precoding~\cite{Zhao_Guo_Yang_2024}. In~\cite{Guo_Wen_Jin_Li_2020, Lin_Lee_Ding_2024}, branched NN designs were introduced to support dynamic compression rates in CSI feedback. To support online adaptation to dynamic system configurations, the multi-user precoding scheme in~\cite{Zhang_Gao_Wen_Cheng_Yang_2025} updates only the user-specific model for each newly arriving user, enabling the system to dynamically handle user arrivals and departures. Despite these advances, we observe that most existing approaches are designed to scale only along a single dimension of the system configuration space. In end-to-end FDD precoding, the scalability challenge is amplified by a high-dimensional configuration space (e.g., number of antennas, number of users, pilot length, and feedback budget), which can render the above solutions ineffective. Moreover, knowledge sharing and reuse across different configurations is rarely considered, leading to redundant training and limited generalization to unseen settings.


In fact, precoding tasks under different system configurations are not independent but exhibit intrinsic correlations. This observation suggests that exposing a model to multi-dimensional heterogeneous configurations during training enables it to learn generalizable precoding knowledge that is decoupled from any specific configuration.  Meanwhile, emerging \emph{wireless foundation models}~\cite{Liu_Gao_Liu_Cheng_Yang_2025, Liu_Gao_Liu_Cheng_Yang_2025_2, cheng2025foundation} show that large-scale pretraining across diverse channel distributions enables robust generalization to unseen wireless environments.  Motivated by this, we propose a \underline{wi}reless \underline{fo}undation model for FDD \underline{e}nd-to-end precoding (\textbf{WiFo-E}) that leverages shared representation learning to support scalability across a wide range of system configurations. In contrast to prior studies, our approach provides a more general solution for enhancing scalability across different system configurations: WiFo-E maps task-specific representations under heterogeneous configurations into a unified feature space, enabling efficient scaling to multi-dimensional heterogeneous configurations in wireless networks. Specifically, during the pretraining stage, the shared backbone adopts multi-task learning to acquire wireless knowledge decoupled from any specific configuration, which facilitates knowledge reuse and sharing across diverse system configurations. During the fine-tuning stage, the pretrained backbone with the learned general knowledge enables rapid adaptation to new configurations with minimal computational overhead. Training such a network, however, is itself challenging, as a fully dense backbone easily incurs task interference across configurations. To mitigate this, we adopt sparse Mixture-of-Experts (MoE) Transformer~\cite{Lepikhin_Lee_Xu_Chen_Firat_Huang_Krikun_Shazeer_Chen_2021} as the backbone, which performs conditional computation and activates only the relevant parts of the model by dynamically routing each input to a small subset of specialized experts. This mechanism reduces conflict among configurations and enables the backbone to capture shared knowledge while separating configuration-specific parts, thereby improving knowledge reuse and overall efficiency.

The main contributions of this paper are summarized as follows:
\begin{enumerate}
    \item \textbf{We formulate the model scalability problem of end-to-end FDD precoding under heterogeneous wireless system configurations as a multi-task learning problem and propose the training framework of WiFo-E.} Unlike prior works, WiFo-E enables knowledge sharing and reuse via multi-task pretraining and allows a single model to scale across multi-dimensional configurations efficiently.

    \item \textbf{We introduce the sparse MoE Transformer as the backbone to mitigate task interference during multi-task pretraining.} By dynamically activating only the relevant experts, the MoE backbone effectively separates configuration-specific patterns while capturing shared structural knowledge, thereby enhancing the backbone's training efficiency and scalability across heterogeneous configurations.
    
    \item \textbf{We conduct extensive simulations to validate the effectiveness and efficiency of WiFo-E.} The results demonstrate that joint training across heterogeneous systems outperforms training each configuration separately. Moreover, the pretrained WiFo-E exhibits rapid adaptation to unseen system configurations during fine-tuning. The sparse MoE architecture also improves multi-task learning performance and offers superior computational efficiency compared to its dense alternatives.
    
\end{enumerate}

The remainder of this paper is organized as follows. Section II introduces the system model for end-to-end precoding. Section III details the heterogeneous pretraining and fine-tuning procedures and the model architecture. Section IV describes the simulation setup and baseline methods. Section V presents and discusses the experimental results. Finally, Section VI concludes the paper and discusses future work.

\textit{Notation:} Scalars are denoted by italic letters (e.g., $a$), vectors by bold lowercase letters (e.g., $\mathbf{a}$), and matrices by bold uppercase letters (e.g., $\mathbf{A}$). The sets of real and complex numbers are represented by $\mathbb{R}$ and $\mathbb{C}$, respectively. Script letters in mathcal font (e.g., $\mathcal{A}$) are used to denote sets or functions, according to their specific definitions within the context. The transpose, conjugate transpose (Hermitian), inverse, absolute value, and $\ell_2$-norm of a matrix or vector are denoted by $(\cdot)^{\rm T}$, $(\cdot)^{\rm H}$, $(\cdot)^{-1}$, $|\cdot|$, and $\|\cdot\|_2$, respectively. The expectation operator is represented as $\mathbb{E}\{\cdot\}$.

\section{System Model and Problem Formulation}

This section introduces the system model of an FDD multi-user MIMO system and formulates an end-to-end precoding problem that jointly designs pilot signaling, channel feedback, and precoding under heterogeneous configurations.

\subsection{Signal Model in wireless networks}
We consider a FDD multi-user MIMO system in a wireless network~\cite{hu2021distributed}, where the BS is equipped with $N_{\rm t}$ antennas and serves $K$ single-antenna users. We assume that the BS employs linear precoding to transmit the data symbol vector $\mathbf{s} \in \mathbb{C}^{K}$ to the users. The precoded signal vector $\mathbf{x} \in \mathbb{C}^{N_{\rm t}}$ can be expressed as:
\begin{equation}
\mathbf{x} = \mathbf{V} \mathbf{s},
\end{equation}
where $\mathbf{V} \in \mathbb{C}^{N_{\rm t} \times K}$ is the precoding matrix satisfying the total power constraint, i.e., ${\rm Tr}(\mathbf{V} \mathbf{V}^{\rm H}) \leq P$. Specifically, the $k$-th column $\mathbf{v}_k$ of $\mathbf{V}$ represents the precoding vector assigned to user $k$, and the $k$-th element $s_k$ of $\mathbf{s}$ denotes the data symbol intended for user $k$. The precoded signal is then broadcast to the users through downlink channels $\mathbf{h}_k \in \mathbb{C}^{N_{\rm t}}$. The received signal at the $k$-th user in the data transmission phase can be expressed as:
\begin{equation}
y_k = \mathbf{h}^{\rm H}_k \mathbf{v}_k s_k + \sum_{j \neq k} \mathbf{h}^{\rm H}_k \mathbf{v}_j s_j + n_k,
\end{equation}
where $\mathbf{h}_k \in \mathbb{C}^{N{\rm t}}$ is the downlink channel vector from the BS to user $k$, and $n_k$ denotes the additive white Gaussian noise (AWGN) at user $k$, modeled as a zero-mean complex Gaussian random variable with variance $\sigma^2$. The first term represents the desired signal, while the second term represents the interference from other users. According to the received signal model, the achievable rate of the $k$-th user can be expressed as:
\begin{equation}
R_k = \log_2\left(1 + \frac{|\mathbf{h}^{\rm H}_k \mathbf{v}_k|^2}{\sum_{j \neq k} |\mathbf{h}^{\rm H}_k \mathbf{v}_j|^2 + \sigma^2}\right).
\end{equation}

In this paper, the goal of precoding design is to maximize the sum rate of all users:
\begin{equation}
\label{eq:sum_rate}
\max_{\mathbf{V}} \quad \sum_{k=1}^K R_k \qquad\qquad\qquad\qquad\qquad
\end{equation}

\addtocounter{equation}{-1}
\vspace{-10pt}

\begin{subequations}
\label{eq:sum_rate_problem}
\begin{align}
\quad\text{s.t.}\quad 
\label{eq:sum_rate_precoder_power}
& \mathrm{Tr}(\mathbf{V}\mathbf{V}^{\mathrm H}) \le P, \\
\label{eq:sum_rate_pilot_power}
& \|\mathbf{x}_\ell\|_2^2 \le E_{\mathrm s}, 
\quad \forall \ell \in \{1,2,\ldots,L\}.
\end{align}
\end{subequations}

\subsection{End-to-End Precoder Design}
\label{sec:End-to-End Precoder Design}

To design the optimal precoding matrix, instantaneous CSI must be aggregated to solve the optimization problem in \eqref{eq:sum_rate}. Modular designs of channel estimation, feedback, and precoding suffer from inconsistent training objectives and inevitable error propagation across stages, leading to degraded overall system performance. Therefore, we adopt an end-to-end FDD precoding framework to jointly optimize the entire process~\cite{Guo_Chen_Ai_Li_2025, Jang_Lee_Kim_Lee_2022}. Specifically, prior to the data transmission phase, the BS initiates a downlink training phase by transmitting a sequence of pilot signals $\mathbf{X} \in \mathbb{C}^{N_{\rm t} \times L}$ over $L$ time slots. During this phase, each user observes a noisy version of the pilot signal after passing through their respective downlink channel. The received pilot signal $\mathbf{y}_k \in \mathbb{C}^{1\times L}$ at user $k$ is given by
\begin{equation}
\label{eq:received_pilot}
\mathbf{y}_k = \mathbf{h}_k^{\rm H} \mathbf{X} + \mathbf{z}_k,
\end{equation}
where $\mathbf{z}_k \sim \mathcal{CN}(0, \sigma^2 \mathbf{I})$ is the AWGN at user $k$. The pilot sequence $\mathbf{x}_\ell$ used at the $\ell$-th time slot (i.e., the $\ell$-th column of $\mathbf{X}$) is subject to a per-symbol power constraint $\|\mathbf{x}_\ell\|_2^2 \leq E_{\rm s}, \quad \forall \ell \in \{1, 2, \ldots, L\}$. After receiving $\mathbf{y}_k$, each user applies a limited feedback scheme $\mathcal{F}_k: \mathbb{C}^{L} \to \{ \pm 1 \}^B$ to compress the observed signals into binary feedback messages:
\begin{equation}
\label{eq:feedback}
\mathbf{q}_k = \mathcal{F}_k(\mathbf{y}_k).
\end{equation}

The compressed messages $\mathbf{q}_k$ are subsequently transmitted via uplink channels, where the limited feedback payload ensures error-free reconstruction at the BS. The BS then collects the feedback messages from all $K$ users into a matrix $\mathbf{Q} \in \{ \pm 1 \}^{B \times K}$ defined as:
\begin{equation}
\label{eq:feedback_matrix}
\mathbf{Q} = [\mathbf{q}_1, \mathbf{q}_2, \ldots, \mathbf{q}_K],
\end{equation}
and uses it to compute the precoding matrix via a mapping function \( \mathcal{G}: \{ \pm 1 \}^{B \times K} \to \mathbb{C}^{N_{\rm t} \times K} \):
\begin{equation}
\label{eq:precoding}
\mathbf{V} = \mathcal{G}(\mathbf{Q}).
\end{equation}

Under this framework, the original objective in equation~\eqref{eq:sum_rate} can be reformulated as a joint optimization problem over the training pilots $\mathbf{X}$, the user-side feedback functions $\{\mathcal{F}_k(\cdot)\}_{k=1}^K$, and the BS-side precoding function $\mathcal{G}(\cdot)$, i.e., 
\begin{equation}
  \label{eq:opt_sum_rate}
  \max_{\mathbf{X}, \{\mathcal{F}_k\}, \mathcal{G}} \sum_{k=1}^K \log_2\left(1 + \frac{|\mathbf{h}_k^{\rm H} \mathbf{v}_k|^2}{\sum_{j \neq k} |\mathbf{h}_k^{\rm H} \mathbf{v}_j|^2 + \sigma^2}\right),
\end{equation}
subject to the power constraints in equations~\eqref{eq:sum_rate_precoder_power}, \eqref{eq:sum_rate_pilot_power} and and the system relations in equations~\eqref{eq:received_pilot}, \eqref{eq:feedback}, \eqref{eq:feedback_matrix}, and \eqref{eq:precoding}.

The system thereby learns an end-to-end mapping from pilot design, through user-side feedback, to BS-side precoding, with the objective of maximizing the achievable downlink sum rate.

\begin{figure*}[!ht]
  \centering
  \includegraphics[width=0.90\linewidth]{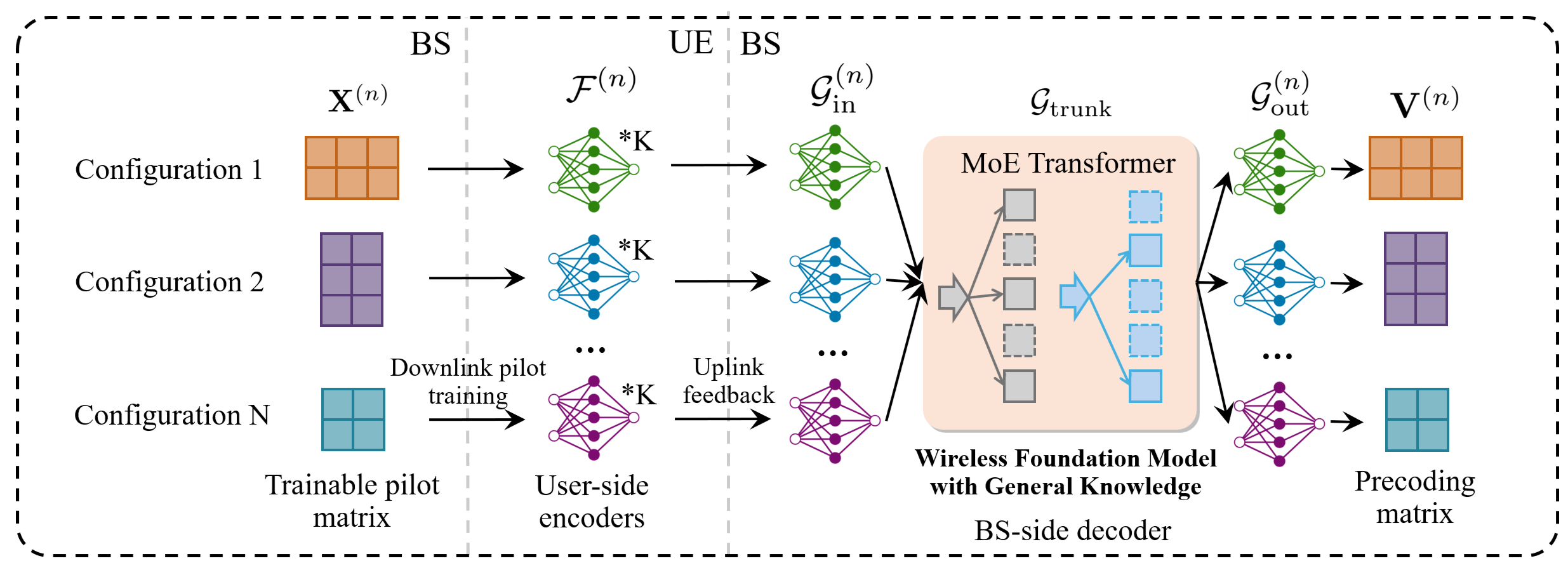}
  \caption{An overview of proposed multi-task pretraining framework.}
  \label{fig:framework}
\end{figure*}

\subsection{Problem Formulation}
In practical wireless networks, system configurations are heterogeneous due to variations in devices, environments, and operational conditions. For instance, the number of antennas differs across base stations, the pilot length and feedback budget are dynamically adjusted with channel quality and resource constraints, and the number of active users varies randomly as users join or leave the network. To address such heterogeneity, we formulate the problem in~\eqref{eq:opt_sum_rate} under multiple configurations as a multi-task learning problem. Let there be \(N\) heterogeneous tasks, each corresponding to a distinct configuration 
\begin{equation}
\mathcal{C}_{(n)}=\{L_{(n)},N_{{\rm t}, (n)},K_{(n)},B_{(n)}\},
\end{equation}
and a channel dataset 
\begin{equation}
\mathcal{D}_{(n)}=\{\mathbf{H}_{i, (n)}\}_{i=1}^{S_{(n)}},
\end{equation} 
which contains \(S_{(n)}\) channel realizations. The heterogeneity spans the pilot length \(L_{(n)}\), the number of transmit antennas \(N_{{\rm t}, (n)}\), the number of users \(K_{(n)}\), and the feedback budget \(B_{(n)}\). The superscript \((n)\) indicates that the variables are defined under the \(n\)-th task. For a given configuration \(\mathcal{C}_{(n)}\), the end-to-end precoding pipeline is specified by the pilot matrix \(\mathbf{X}_{(n)}\in\mathbb{C}^{N_{{\rm t}, (n)}\times L_{(n)}}\), user-side feedback mappings \(\{\mathcal{F}_{k,(n)}:\mathbb{C}^{L_{(n)}}\!\to\!\{\pm1\}^{B_{(n)}}\}_{k=1}^{K_{(n)}}\), and a BS-side precoder mapping \(\mathcal{G}_{(n)}:\{\pm1\}^{B_{(n)}\times K_{(n)}}\!\to\!\mathbb{C}^{N_{{\rm t}, (n)}\times K_{(n)}}\).

Rather than optimizing each task in isolation, we consider a joint multi-task design that encapsulates the network construction and training process, defined by  
\begin{equation}
\label{c11}
 \mathcal{C}_{(n)}\mapsto\big(\mathbf{X}_{(n)},\,\{\mathcal{F}_{k, (n)}\},\,\mathcal{G}_{(n)}\big),
\end{equation}
which represents the design framework and learning mechanism that determine the pilot, feedback, and precoding modules corresponding to each system configuration \(\mathcal{C}_{(n)}\). Under this framework, the resulting precoding matrix \(\mathbf{V}_{(n)}\) is obtained as the learned outcome of the end-to-end network trained under configuration \(\mathcal{C}_{(n)}\). The optimization problem is finally summarized as
\begin{equation}
\label{eq:hetero_objective_compact}
\max_{\left\{\mathbf{X}_{(n)},\,\{\mathcal{F}_{k, (n)}\},\,\mathcal{G}_{(n)}\right\}}
\ \sum_{n=1}^{N} w_n\ \mathbb{E}_{\mathbf{H}\sim\mathcal{D}_{(n)}}\!\left[\ \sum_{k=1}^{K_{(n)}} R_{k, (n)}\ \right]
\end{equation}

\addtocounter{equation}{-1}
\vspace{-10pt}

\begin{subequations}
\label{eq:hetero_constraints_compact}
\begin{align}
\text{s.t.}\quad &\mathbf{y}_{k, (n)} = {\mathbf{h}^{\rm H}_{k, (n)}}\, \mathbf{X}_{(n)} + \mathbf{z}_k, \label{eq:c_observation}\\
& \mathbf{q}_{k, (n)} = \mathcal{F}_{k, (n)}\!\left(\mathbf{y}_{k, (n)}\right), \label{eq:c_feedback}\\
& \mathbf{V}_{(n)} = \mathcal{G}_{(n)}\!\left([\mathbf{q}_{1, (n)},\ldots,\mathbf{q}_{K_{(n)},(n)}]\right). \label{eq:c_precoding}\\ & \|\mathbf{x}_{\ell, (n)}\|_2^2 \le E_{\rm s}, \label{eq:c_pilot_energy}
\\ & \mathrm{Tr}\big(\mathbf{V}_{(n)}{\mathbf{V}^{\rm H}_{(n)}\big)} \le P_{(n)}, \label{eq:c_power}
\end{align}
\end{subequations}
for every task \(n\) and user \(k\). Here \(w_n\!\ge\!0\) are aggregation weights (e.g., uniform or proportional to \(|\mathcal{D}_{(n)}|\)). This formulation explicitly captures the requirement to maximize downlink sum rate across multi-dimensional heterogeneous configurations--constrained by per-task pilot length, antenna count, user count, and feedback budget.

\section{Wireless Foundation Model for End-to-End Precoding}

In this section, we introduce the multi-task pretraining, fine-tuning strategy for unseen systems, and the shared backbone architecture of the proposed WiFo-E framework.


\subsection{Pretraining via Multi-Task Learning}


To efficiently implement the design in equation~\eqref{c11}, we adopt a multi-task learning paradigm to pretrain a foundation model, as illustrated in Figure~\ref{fig:framework}. In deployment, recent standardization efforts on AI/ML for the radio access network (RAN) and the NR air interface generally advocate placing the primary computational burden on the network side (e.g., the BS/gNB), while retaining a certain degree of flexibility at the user equipment (UE) through lightweight neural modules~\cite{3gpp_tr38843,etsi_ts_103_983}. This design naturally leads to a BS-centric architecture with minimized UE complexity. In addition, to avoid the storage and maintenance overhead associated with deploying per-user models at the users, we reuse a single user-side parameter set across users. Accordingly, for each system configuration \(\mathcal{C}_{(n)}\), the BS-side model is decomposed into three components, namely \(\mathcal{G}_{{\rm in},(n)}\), \(\mathcal{G}_{\rm trunk}\), and \(\mathcal{G}_{{\rm out},(n)}\). The model \(\mathcal{G}_{\rm trunk}\) acts as a shared backbone across all tasks, whereas the modules \(\mathcal{G}_{{\rm in},(n)}\) and \(\mathcal{G}_{{\rm out},(n)}\) serve as task-specific branches that adapt the model to the particular task \(n\). Because the feedback overhead and the dimensions of the precoding matrices vary across tasks, the input dimension of \(\mathcal{G}_{{\rm in},(n)}\) and the output dimension of \(\mathcal{G}_{{\rm out},(n)}\) are generally different. Consequently, each \(\mathcal{G}_{{\rm in},(n)}\) first maps the binary feedback into a common feature space of fixed dimension; after passing through the shared backbone, \(\mathcal{G}_{{\rm out},(n)}\) reconstructs the precoding matrix from this common feature representation. In addition, the ordering of user feedback should not affect the precoding vector assigned to each user. We therefore employ user-side networks \(\mathcal{F}_{(n)}\) with shared parameters for all users under task \(n\), instead of learning a separate network \(\mathcal{F}_{k,(n)}\) for each user.

For each precoding task \(n\), we train a neural network model parameterized by \(\Theta_{(n)} = \{ \Theta_{\rm trunk}, \Theta_{{\rm specific},(n)} \}\), where \(\Theta_{\rm trunk}\) denotes the parameters of the shared backbone \(\mathcal{G}_{\rm trunk}\), and \(\Theta_{{\rm specific},(n)}\) corresponds to the task-specific parameters of \(\mathcal{F}_{(n)}\), \(\mathcal{G}_{{\rm in}, (n)}\), and \(\mathcal{G}_{{\rm out}, (n)}\) for task \(n\). The complete set of trainable parameters in the model is given by
\begin{equation}
\Theta =\Theta_{\rm trunk} \cup \bigcup_{n=1}^N \Theta_{{\rm specific},(n)}   .
\end{equation}
To optimize the entire model \(\Theta\) during the pretraining stage, each task is associated with a task-specific loss function \(\mathcal{L}_{(n)}(\mathcal{D}_{(n)}; \Theta_{(n)})\), defined as the negative expected sum rate over the dataset \(\mathcal{D}_{(n)}\), i.e.,
\begin{equation}
\label{eq:loss}
\mathcal{L}_{(n)}(\mathcal{D}_{(n)}; \Theta_{(n)}) = - \mathbb{E}_{\mathbf{H}_{(n)} \sim \mathcal{D}_{(n)}} \left[ \sum_{k=1}^{K_{(n)}} R_{k,(n)} \right],
\end{equation}
where $R_{k, (n)}$ denotes the achievable rate of user $k$ under configuration $\mathcal{C}_{(n)}$, following the expression given in equation~\eqref{eq:sum_rate}. The overall loss function \(\mathcal{L}_{\rm MTL}\), representing a weighted sum of task-specific losses across all tasks, is given by
\begin{equation}
\mathcal{L}_{\rm MTL} = \sum_{n=1}^N \lambda_{(n)} \mathcal{L}_{(n)}(\mathcal{D}_{(n)}; \Theta_{(n)}),
\end{equation}
where \(\lambda_{(n)} \in \mathbb{R}^{+}\) denotes the weight assigned to task \(n\). The overall pretraining process is summarized in Algorithm~\ref{alg:mtl-pretrain}.

A key motivation for introducing a shared backbone is that different tasks exhibit some similarity, such that multi-task learning can promote transfer and reuse of knowledge. Under the proposed framework, knowledge that is common across tasks and agnostic to specific system parameters is naturally captured in the shared backbone, whereas task-specific knowledge is captured by the task-specific branches. Nevertheless, jointly training multiple tasks on a single backbone may induce task interference, which can cause the performance of some tasks to degrade compared with training independent models. To mitigate this, Subsection~\ref{model_arch} proposes using a sparse MoE architecture instead of a dense backbone. The MoE performs conditional computation by routing each input to a small subset of experts, which reduces interference among tasks and allows the backbone to learn shared representations while separating task-specific behaviors~\cite{Shazeer_Mirhoseini_Maziarz_Davis_Le_Hinton_Dean_2017, Chen_Shen_Ding_Chen_Zhao_Learned_Miller_Gan_2023}.


\begin{algorithm}[!ht]
\caption{Multi-Task Learning in Pretraining Stage}
\label{alg:mtl-pretrain}
\DontPrintSemicolon
\SetKwInput{KwIn}{Input}
\SetKwInput{KwOut}{Output}
\KwIn{Task set $\mathcal{T}=\{\mathcal{T}_{(n)}\}_{n=1}^{N}$ with $\mathcal{T}_{(n)}=\{\mathcal{D}_{(n)},\mathcal{C}_{(n)}\}$; task weights $\{\lambda_{(n)}\}_{n=1}^N$; epochs $E$; mini-batch sizes $\{M_{(n)}\}$; optimizer $\mathsf{Opt}$; learning rate $\eta$.}
\KwOut{Optimized parameters $\Theta=\Theta_{\rm trunk}\cup\bigcup_{n=1}^N\Theta_{{\rm specific}, (n)}$.}
\BlankLine
\textbf{Initialize} $\Theta_{\rm trunk}$ and $\{\Theta_{{\rm specific}, (n)}\}_{n=1}^{N}$, including normalized pilots $\{\mathbf{X}_{(n)}\}$, user models $\{\mathcal{F}_{(n)}\}$, task-specific BS models $\{\mathcal{G}_{{\rm in},(n)},\mathcal{G}_{{\rm out},(n)}\}$, and the shared backbone $\mathcal{G}_{\rm trunk}$.\;
\For{$e\gets 1$ \KwTo $E$}{
  $\mathcal{L}_{\rm MTL}\gets 0$ 
  \For{$n\gets 1$ \KwTo $N$}{
    Sample mini-batch $\{\mathbf{H}_{i, (n)}\}_{i=1}^{M_{(n)}}\subset\mathcal{D}_{(n)}$.\;
    \For{$i\gets 1$ \KwTo $M_{(n)}$}{
      $\mathbf{H}_{(n)}\gets \mathbf{H}_{i, (n)}$.\;
      Transmit pilot $\mathbf{X}_{(n)}$ and get $\{\mathbf{y}_{k,(n)}\}_{k=1}^{K_{(n)}}$.\;
      $\mathbf{q}_{k,(n)}\gets \mathcal{F}_{(n)}(\mathbf{y}_{k,(n)})$ for $k=1,\dots,K_{(n)}$; form $\mathbf{Q}_{(n)}=[\mathbf{q}_{1, (n)},\ldots,\mathbf{q}_{K_{(n)},(n)}]$.\;
      $\mathbf{Z}_{(n)}\gets \mathcal{G}_{{\rm in},(n)}(\mathbf{Q}_{(n)})$; $\tilde{\mathbf{Z}}_{(n)}\gets \mathcal{G}_{\rm trunk}(\mathbf{Z}_{(n)})$; $\mathbf{V}_{(n)}\gets \mathcal{G}_{{\rm out},(n)}(\tilde{\mathbf{Z}}_{(n)})$.\;
      $\ell_{(n)}(\mathbf{H}_{(n)};\Theta_{(n)})\gets -\sum_{k=1}^{K_{(n)}} R_{k,(n)}$.\;
    }
    $\mathcal{L}_{(n)}\gets \frac{1}{M_{(n)}}\sum_{i=1}^{M_{(n)}} \ell_{(n)}(\mathbf{H}_{i, (n)};\Theta_{(n)})$.\;
    $\mathcal{L}_{\rm MTL}\gets \mathcal{L}_{\rm MTL}+\lambda_{(n)}\mathcal{L}_{(n)}$.\;
  }
  Compute $\nabla_{\Theta}\mathcal{L}_{\rm MTL}$ by backpropagation.\;
  $\Theta \gets \mathsf{Opt}(\Theta,\nabla_{\Theta}\mathcal{L}_{\rm MTL},\eta)$.\;
}
\Return{$\Theta$}\;
\end{algorithm}

\subsection{Fine-Tuning for Unseen Tasks}

In deployment, the model may need to handle precoding tasks under previously unseen configurations in the pretraining stage. As configurations may differ in the number of antennas, the number of users, pilot length, and feedback overhead, directly applying the pretrained model is generally infeasible. In this case, a fine-tuning procedure is conducted to adapt the model to the new task \(m\). Specifically, the pretrained parameters \(\Theta_{\rm trunk}\) of the shared backbone in \(\mathcal{G}_{\rm trunk}\) are retained and frozen, while a new set of task-specific models \(\mathcal{F}^{(m)}\), \(\mathcal{G}_{\rm in}^{(m)}\), and \(\mathcal{G}_{\rm out}^{(m)}\) is initialized with trainable parameters \(\Theta^{(m)}_{\rm specific}\). The resulting model for task \(m\) is thus parameterized by \(\Theta^{(m)} = \{\Theta_{\rm trunk}, \Theta^{(m)}_{\rm specific}\}\) and the trainable part is optimized on the dataset \(\mathcal{D}^{(m)}\) using the same loss form as in pretraining:  
\begin{equation}
\label{eq:fine-tuning}
\mathcal{L}_{\rm fine-tuning} = - \mathbb{E}_{\mathbf{H}^{(m)} \sim \mathcal{D}^{(m)}} \left[ \sum_{k=1}^{K^{(m)}} R_k^{(m)} \right],
\end{equation}
where \(K^{(m)}\) denotes the number of users under configuration \(\mathcal{C}^{(m)}\). This design is motivated by the observation that end-to-end precoding tasks under different configurations are often correlated. The shared backbone is expected to learn the general knowledge that can be transferred across tasks, while the task-specific branches capture the differences associated with each task. In this transfer learning setting, the shared knowledge obtained during multi-task pretraining is reused to accelerate adaptation to a new task and to reduce the amount of training data required, while the frozen backbone helps preserve the general representations learned across all tasks in the pretraining stage. The overall fine-tuning process is summarized in Algorithm~\ref{alg:fine-tune}.

\begin{algorithm}[t]
\caption{Fine-Tuning for Unseen Task $m$}
\label{alg:fine-tune}
\DontPrintSemicolon
\SetKwInput{KwIn}{Input}
\SetKwInput{KwOut}{Output}
\KwIn{Pretrained and \textbf{frozen} backbone $\Theta_{\rm trunk}$; unseen system data $\mathcal{D}^{(m)}$ and configuration $\mathcal{C}^{(m)}$; epochs $E$; mini-batch size $M$; optimizer $\mathsf{Opt}$; learning rate $\eta$.}
\KwOut{Fine-tuned task-specific parameters $\Theta_{\rm specific}^{(m)}$.}
\BlankLine
\textbf{Initialize} new task-specific model $\Theta_{\rm specific}^{(m)}$ (normalized pilot $\mathbf{X}^{(m)}$, user encoder $\mathcal{F}^{(m)}$, BS maps $\mathcal{G}^{(m)}_{\rm in},\,\mathcal{G}^{(m)}_{\rm out}$).\;
Freeze all parameters in $\Theta_{\rm trunk}$; enable gradients only for $\Theta_{\rm specific}^{(m)}$.\;
\For{$e\gets 1$ \KwTo $E$}{
  \For{each $\{\mathbf{H}^{(m)}_i\}_{i=1}^{M}\subset\mathcal{D}^{(m)}$}{
    $\mathcal{L}_{\rm fine}\gets 0$.\;
    \For{$i\gets 1$ \KwTo $M$}{
      $\mathbf{H}^{(m)}\gets \mathbf{H}^{(m)}_i$.\;
      Transmit pilot $\mathbf{X}^{(m)}$ and get $\{\mathbf{y}^{(m)}_{k}\}_{k=1}^{K^{(m)}}$.\;
      $\mathbf{Q}^{(m)}_i\gets[\mathcal{F}^{(m)}(\mathbf{y}^{(m)}_{1}),\ldots,\mathcal{F}^{(m)}(\mathbf{y}^{(m)}_{K^{(m)}})]$.\;
      $\mathbf{Z}^{(m)}_i\gets\mathcal{G}^{(m)}_{\rm in}(\mathbf{Q}^{(m)}_i)$;\quad
      $\tilde{\mathbf{Z}}^{(m)}_i\gets\mathcal{G}_{\rm trunk}(\mathbf{Z}^{(m)}_i)$;\quad
      $\mathbf{V}^{(m)}_i\gets\mathcal{G}^{(m)}_{\rm out}(\tilde{\mathbf{Z}}^{(m)}_i)$.\;
      $\mathcal{L}_{\rm fine}\gets\mathcal{L}_{\rm fine}
      -\frac{1}{M}\sum_{k=1}^{K^{(m)}} R_k^{(m)}$.\;
    }
    Compute $\nabla_{\Theta_{\rm specific}^{(m)}} \mathcal{L}_{\rm fine}$.\;
    $\Theta_{\rm specific}^{(m)} \gets \mathsf{Opt}\!\left(\Theta_{\rm specific}^{(m)},\,\nabla_{\Theta_{\rm specific}^{(m)}} \mathcal{L}_{\rm fine},\,\eta\right)$.\;
  }
}
\Return{$\Theta_{\rm specific}^{(m)}$ (with $\Theta_{\rm trunk}$ unchanged).}
\end{algorithm}

\subsection{Model Architecture}
\label{model_arch}

The proposed WiFo-E framework adopts a unified end-to-end  architecture that jointly learns the pilot design, user-side feedback encoding, and BS-side precoding across heterogeneous configurations. The overall model follows an encoder--decoder paradigm, where each user independently encodes its received pilot signals into a compact feedback representation, and the BS aggregates the feedback from all users to generate the corresponding precoding matrix. Next, we introduce the network components on the user and BS sides. For brevity, the task index superscript $(\cdot)_{(n)}$ is omitted in this subsection.

\subsubsection{User-side encoders for channel feedback}
On the user side, the encoder $\mathcal{F}$ uses a lightweight multi-layer perceptron (MLP) consisting of $L$ layers to map the received pilot signal $\mathbf{y}_k \in \mathbb{C}^{L}$ to a binary feedback message $\mathbf{q}_k \in \{\pm 1\}^{B}$. The transformation is given by nested functions:
\begin{subequations}
\label{eq:mlp}
\begin{align}
\mathbf{h}[0] &= \mathbf{y}_k, \label{eq:mlp_a} \\
\mathbf{h}[\ell] &= \sigma\!\left( \mathbf{W}[\ell] \mathbf{h}[\ell-1] + \mathbf{b}[\ell] \right), \label{eq:mlp_b}
\end{align}
\end{subequations}
where $\mathbf{W}[\ell]$ and $\mathbf{b}[\ell]$ denote the weight matrix and bias vector of the $\ell$-th layer, respectively, for all $\ell=1,2,\ldots,L$, and $\sigma(\cdot)$ represents the activation function (e.g., ReLU). The final output of the MLP, $\mathbf{h}[L]$, is further quantized to produce the binary feedback message $\mathbf{q}_k \in \{\pm 1\}^{B}$. These binary feedback vectors $\mathbf{q}_k$ are subsequently transmitted to the BS via the uplink channels.

\subsubsection{BS-side decoder for precoding}

The BS aggregates compact user feedback into
\begin{equation}
\label{eq:bs_Q_agg}
\mathbf{Q}=\big[\mathbf{q}_1,\ldots,\mathbf{q}_{K}\big]\in\{\pm 1\}^{B\times K}.
\end{equation}
We project $\mathbf{Q}$, whose dimension depends on the configuration, to a unified space of width $d$ and arrange $K$ user tokens as
\begin{equation} \label{eq:bs_proj}
\mathbf{Z}[0] = \mathbf{Q}^{\top} \mathbf{W}_{\rm in} + \mathbf{1}_K \mathbf{b}_{\rm in}^{\top},
\end{equation}
where $\mathbf{W}_{\rm in} \in \mathbb{R}^{B \times d}$ is a learnable projection matrix, $\mathbf{b}_{\rm in} \in \mathbb{R}^d$ is a bias vector, and $\mathbf{1}_K \in \mathbb{R}^{K\times 1}$ is an all-ones column vector. Positional encoding is omitted in our model because permutation equivariance across users ensures that reordering users does not affect model performance.

\begin{figure}[!t]
  \centering
  \includegraphics[width=0.95\linewidth]{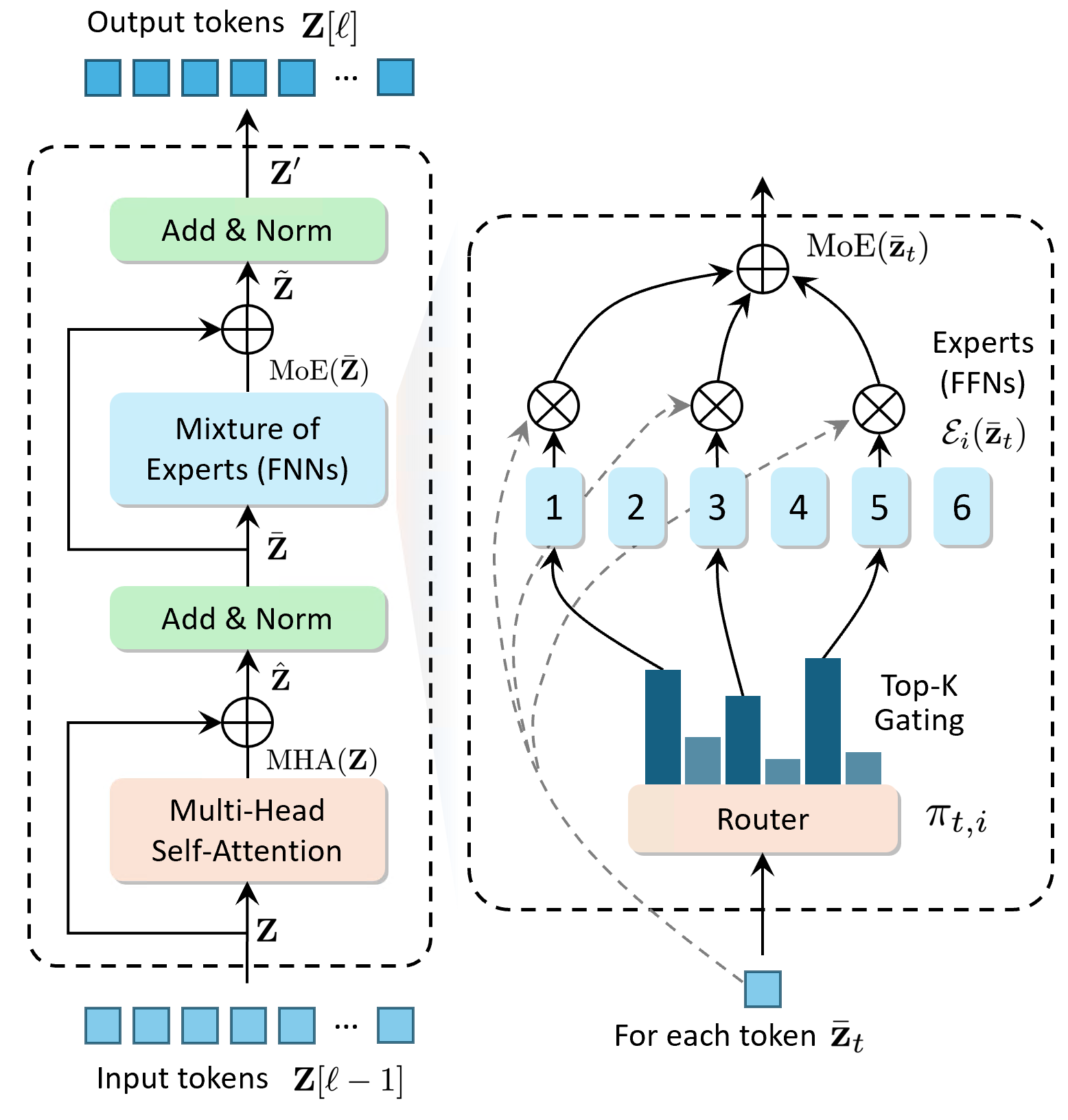}
  \caption{MoE-Transformer block.}
  \label{fig:moe-transformer}
\end{figure}

The MoE-Transformer backbone comprises $L$ stacked blocks that act on the token matrix $\mathbf{Z}[0]$ and produce $\mathbf{Z}[L]$ through nested composition, as shown in Figure~\ref{fig:moe-transformer}. Specifically, letting $\mathcal{B}_\ell(\cdot)$ denote the $\ell$-th block, the recursion is
\begin{equation}
\label{eq:bs_trunk_stack}
\mathbf{Z}[\ell] = \mathcal{B}_\ell\!\big(\mathbf{Z}[\ell-1]\big),\quad \ell=1,\ldots,L.
\end{equation}
Each Transformer block consists of two sublayers, which are described in detail below.

\textbf{(1) Multi-Head Self-Attention (MHSA) Sublayer.}
Each block first applies multi-head self-attention followed by layer normalization with a residual connection:
\begin{subequations}
\label{eq:bs_block_mha}
\begin{align}
\hat{\mathbf{Z}} &= \mathbf{Z} + \mathrm{MHA}(\mathbf{Z}), \label{eq:bs_block_mha_res}\\
\bar{\mathbf{Z}} &= \mathrm{LN}(\hat{\mathbf{Z}}), \label{eq:bs_block_mha_ln}
\end{align}
\end{subequations}
where $\mathbf{Z}$ and $\bar{\mathbf{Z}}$ denote the input to the block and the output of the MHSA sublayer, respectively, both of dimension $K \times d$. Given $H$ attention heads with per-head dimensionality $d_h = d/H$, the MHA output is defined as
\begin{equation}
\label{eq:bs_mha}
\mathrm{MHA}(\mathbf{Z}) = \mathrm{Concat}\big(\mathbf{X}_1,\ldots,\mathbf{X}_H\big)\,\mathbf{W}_{\rm O},
\end{equation}
where $\mathbf{W}_{\rm O} \in \mathbb{R}^{H d_h \times d}$ denotes the output projection matrix. For the $j$-th head, the output $\mathbf{X}_j$ is computed as
\begin{equation}
\label{eq:bs_mha_head}
\mathbf{X}_j = \mathrm{softmax}\!\left(\frac{\mathbf{Q}_j\mathbf{K}_j^{\top}}{\sqrt{d_h}}\right)\mathbf{V}_j,
\end{equation}
where the query, key, and value matrices are given by
\[
\mathbf{Q}_j = \mathbf{Z}\mathbf{W}^{(j)}_{\rm Q},\quad
\mathbf{K}_j = \mathbf{Z}\mathbf{W}^{(j)}_{\rm K},\quad
\mathbf{V}_j = \mathbf{Z}\mathbf{W}^{(j)}_{\rm V},
\]
with learnable matrices $\mathbf{W}^{(j)}_{\rm Q}, \mathbf{W}^{(j)}_{\rm K}, \mathbf{W}^{(j)}_{\rm V} \in \mathbb{R}^{d \times d_h}$.

\textbf{(2) Mixture-of-Experts Feed-Forward Network (MoE-FFN) Sublayer.}
To improve scalability across heterogeneous configurations, we adopt a sparse MoE-FFN with conditional computation via token-wise expert routing.
By routing each token to a small subset of experts, the MoE layer increases the effective model capacity under a limited computation budget, while reducing negative transfer by allowing different configurations to specialize in different experts.
Accordingly, the second sublayer applies a sparse MoE feed-forward network to each token independently, followed by layer normalization with a residual connection:
\begin{subequations}
\label{eq:bs_block_moe}
\begin{align}
\tilde{\mathbf{Z}} &= \bar{\mathbf{Z}} + \mathrm{MoE}(\bar{\mathbf{Z}}), \label{eq:bs_block_moe_res}\\
\mathbf{Z}' &= \mathrm{LN}(\tilde{\mathbf{Z}}), \label{eq:bs_block_moe_ln}
\end{align}
\end{subequations}
where $\mathbf{Z}'\in \mathbb{R}^{K\times d}$ is the block output. For the $t$-th token $\bar{\mathbf{z}}_t$ (the $t$-th row of $\bar{\mathbf{Z}}$), a linear router produces logits
\[
\mathbf{r}_t=\bar{\mathbf{z}}_t\mathbf{W}_{\rm r}+\mathbf{b}_{\rm r},
\]
with $\mathbf{W}_{\rm r}\in\mathbb{R}^{d\times E}$ and $\mathbf{b}_{\rm r}\in\mathbb{R}^E$, from which a top-$k$ expert set $\mathcal{S}_k(\mathbf{r}_t)$ is selected. The mixture weights are computed as
\begin{equation}
\label{eq:bs_pi}
\pi_{t,i} = \frac{\exp(r_{t,i})}{\sum_{j\in \mathcal{S}_k(\mathbf{r}_t)} \exp(r_{t,j})}, \qquad i\in \mathcal{S}_k(\mathbf{r}_t),
\end{equation}
and the MoE output is then formed by the gated aggregation
\begin{equation}
\label{eq:bs_moe}
\mathrm{MoE}(\bar{\mathbf{z}}_t) = \sum_{i\in \mathcal{S}_k(\mathbf{r}_t)} \pi_{t,i}\, \mathcal{E}_i(\bar{\mathbf{z}}_t),
\end{equation}
where each expert is a two-layer position-wise MLP,
\begin{equation}
\label{eq:bs_expert}
\mathcal{E}_i(\mathbf{z}) = \phi\!\big(\mathbf{z}\mathbf{U}_{i,1}+\mathbf{c}_{i,1}\big)\mathbf{U}_{i,2}+\mathbf{c}_{i,2},
\end{equation}
with learnable matrices $\mathbf{U}_{i,1}\in\mathbb{R}^{d\times d_{\rm ff}}$, $\mathbf{U}_{i,2}\in\mathbb{R}^{d_{\rm ff}\times d}$, and $\phi(\cdot)$ a pointwise nonlinearity function.

After the stacked transformation
\begin{equation}
\mathbf{Z}[L] = \big(\mathcal{B}_L \circ \cdots \circ \mathcal{B}_1\big)\!\big(\mathbf{Z}[0]\big),
\end{equation}
the resulting hidden states $\mathbf{Z}[L] \in \mathbb{R}^{K \times d}$ are mapped into a real-valued matrix through a linear output head:
\begin{equation}
\label{eq:bs_out}
\widehat{\mathbf{V}} = \mathbf{Z}[L]\,\mathbf{W}_{\rm out} + \mathbf{1}_K \mathbf{b}_{\rm out}^{\top},
\end{equation}
where $\mathbf{W}_{\rm out} \in \mathbb{R}^{d \times 2N_{\rm t}}$ and $\mathbf{b}_{\rm out} \in \mathbb{R}^{2N_{\rm t}}$. The $k$-th row of $\widehat{\mathbf{V}}$ is interpreted as the concatenation of the real and imaginary parts of the precoding vector for user $k$, which yields a complex-valued vector $\mathbf{v}_k \in \mathbb{C}^{N_{\rm t}}$. Stacking these user-wise precoding vectors as columns gives the precoding matrix
\[
\mathbf{V}_{\rm raw} = [\mathbf{v}_1,\ldots,\mathbf{v}_K] \in \mathbb{C}^{N_{\rm t} \times K},
\]
which is consistent with the system model in Section~II. Finally, the precoding matrix is scaled to meet the total transmit power constraint $P$ via normalization:
\begin{equation}
\label{eq:bs_power_norm}
\mathbf{V}
=
\sqrt{P}\,
\frac{\mathbf{V}_{\rm raw}}{\|\mathbf{V}_{\rm raw}\|_{\mathrm{F}}}.
\end{equation}

\section{Implementation Details}

This section provides the implementation details of the proposed framework, including dataset construction, model training setups, and baseline setups.

\subsection{Dataset Construction}

\begin{table*}[ht]
  \centering
  \caption{\textsc{Dataset configurations for varying MIMO scenarios under 3GPP standards.}}
  \includegraphics[width=0.92\linewidth]{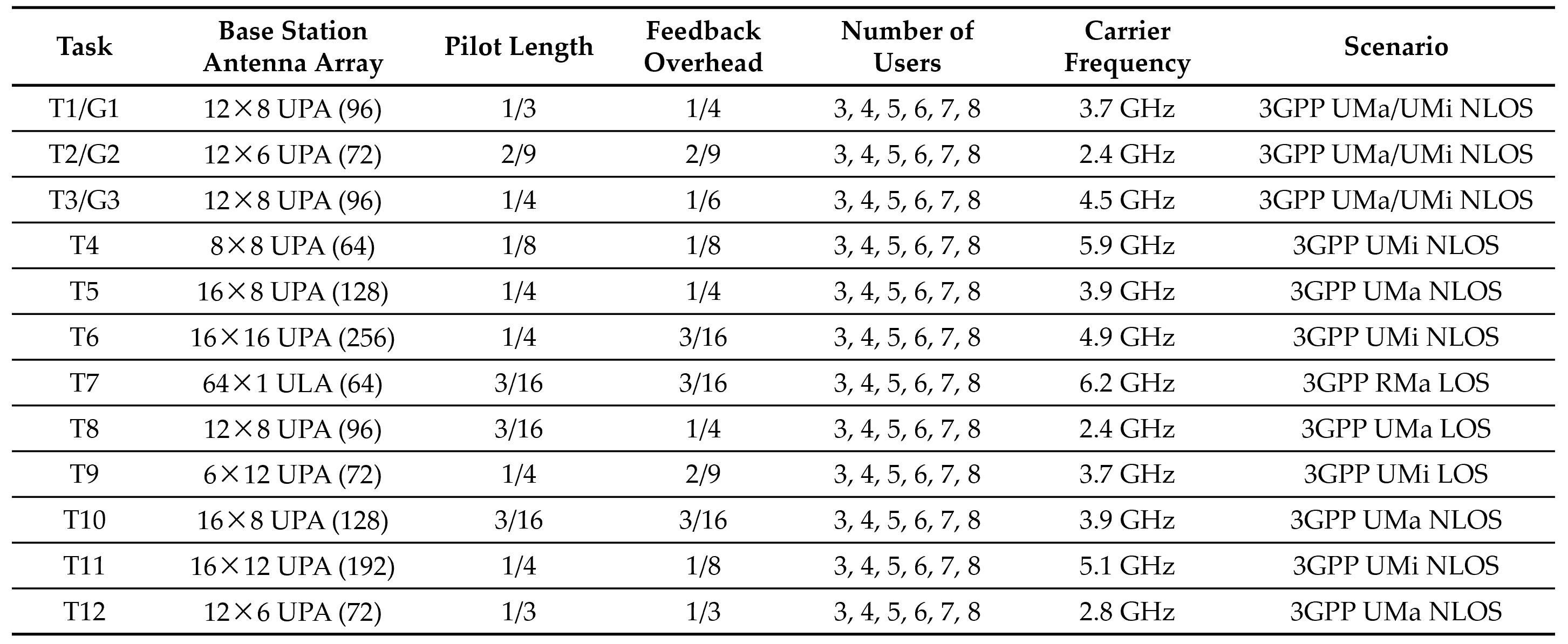}
  \label{tab:dataset_config}
\end{table*}

The simulation experiments were conducted based on channel realizations generated using the Quadriga channel simulator~\cite{jaeckel2014quadriga, abc}. As summarized in Table~\ref{tab:dataset_config}, the dataset encompasses a variety of tasks under heterogeneous system configurations, including different numbers of transmit antennas ($N_{\rm t}$), numbers of users ($K$), pilot sequence lengths ($L$), and feedback overhead ($B$). The values of pilot length and feedback overhead are presented as fractions, representing their ratios with respect to the number of antennas. As an example, for task T9 with $72$ antennas, the pilot length is calculated as $72 \times \tfrac{1}{4} = 18$, while each user is allocated $72 \times \tfrac{2}{9} = 16$ feedback bits. Furthermore, the channel samples span multiple frequency bands ranging from $2.4$~GHz to $6.2$~GHz and encompass diverse propagation environments, including both Line-of-Sight (LOS) and Non-Line-of-Sight (NLOS) conditions in 3GPP Urban Macrocell (UMa), Urban Microcell (UMi), and Rural Macrocell (RMa) scenarios. The overall set of tasks is divided into three groups for different evaluation purposes: \textit{Group~1} (tasks T1 to T9) is used for multi-task pretraining; \textit{Group~2} (tasks T10 to T12) is reserved to assess the model's scalability to tasks that are unseen during the pretraining stage; and \textit{Group~3} (tasks G1 to G3, within the UMa NLOS scenario) is utilized to evaluate the model's domain generalization performance. For each scenario, a total of $300{,}000$ channel samples were generated and subsequently split into training and test sets with a ratio of $7:3$.

\subsection{Model Training Setups}

All neural networks were implemented using the PyTorch framework. Training was conducted with a batch size of $2048$, employing the Adam optimizer~\cite{kingma2014adam} with an initial learning rate of $10^{-3}$. L2 regularization was applied through a weight decay coefficient of $10^{-4}$. A dropout rate of $0.05$ was further introduced to mitigate overfitting during training.

Next, we summarize the model setups used in the proposed framework. On the user side, a lightweight encoder is employed, consisting of $3$ stacked layers with a feed-forward width of $d_{\mathrm{ff}} = 128$. On the BS side, the core architecture adopts a MoE-Transformer backbone composed of $5$ stacked blocks, each with a model width of $d = 256$. Within each block, the MHSA sublayer uses $H = 8$ attention heads, while the MoE-FFN sublayer employs $E = 8$ experts with top-$k = 3$ routing applied independently to each token.

\subsection{Baseline Methods}

To evaluate the effectiveness of the proposed MTL-based scheme, comparisons were conducted against five representative baseline methods:

\textit{1) Zero-Forcing (ZF):} 
ZF precoding is implemented assuming full channel state information at the transmitter (CSIT). The precoding matrix is computed as
\begin{equation}
  \mathbf{V}_{\rm ZF} = \gamma \mathbf{H}^{\rm H} (\mathbf{H} \mathbf{H}^{\rm H})^{-1},
  \label{eq:ZF}
\end{equation}
where $\mathbf{H}$ denotes the downlink channel matrix and $\gamma$ is a normalization factor to satisfy the transmit power constraint~\cite{Spencer_Swindlehurst_Haardt_2004}.

\textit{2) Weighted Minimum Mean Square Error (WMMSE):} 
The WMMSE baseline assumes full CSIT at the BS and computes the precoding matrix via the standard iterative WMMSE optimization procedure that minimizes the weighted sum of users' MSEs~\cite{Shi_Razaviyayn_Luo_He_2011,Christensen_Agarwal_De_Carvalho_Cioffi_2008}.

\textit{3) Single Task Learning (STL):} 
The STL scheme serves as a learning-based baseline, where an independent end-to-end neural network model is trained for each task, and no model parameters are shared across different tasks. In this scheme, both the user-side network $\mathcal{F}_{(n)}(\cdot)$ and the BS-side network $\mathcal{G}_{(n)}(\cdot)$ are trained separately for task \(n\), with the optimization objective of maximizing the system sum rate for each task. The network architecture of STL is identical to that of the proposed MTL framework. In contrast to the proposed MTL scheme, STL does not include a shared backbone nor joint training across tasks, and therefore cannot enable knowledge sharing or transfer across different system configurations. This baseline is used to evaluate the performance gains of multi-task pretraining in terms of scalability and generalization.

\textit{4) Channel Estimation and Precoding (CEP):} The CEP scheme employs a modular design, with channel estimation performed at the user side and precoding at the BS, assuming perfect feedback of the estimated channels. During the training stage, the channel estimation model $\mathcal{F}_{(n)}(\cdot)$ at the user side is optimized using an MMSE loss, while the precoding model $\mathcal{G}_{(n)}(\cdot)$ at the BS side is optimized for maximal sum rate.

\textit{5) Distributed Source Coding (DSC):} 
In~\cite{Sohrabi_Attiah_Yu_2021}, the end-to-end precoding problem is formulated as a distributed source coding (DSC) problem, where different users are modeled as multiple correlated information sources. Each user independently encodes the received pilot signals using a user-side model $\mathcal{F}_{(n)}(\cdot)$, while the BS-side model $\mathcal{G}_{(n)}(\cdot)$ jointly decodes the encoded representations from all users to recover the precoding matrix. Compared with the STL baseline, the key distinction of the DSC method lies in its network realization: both the user-side and BS-side models are implemented using MLPs. Moreover, since the DSC design and training process is limited to a single system configuration, it does not involve parameter sharing or joint training across multiple tasks as well.

To ensure fairness, the backbone architectures of the user-side model $\mathcal{F}_{(n)}(\cdot)$ and the BS-side model $\mathcal{G}_{(n)}(\cdot)$ are kept identical across the MTL, DSC, CEP, and STL schemes, with only minor dimensional differences due to system heterogeneity. For all DL-based methods, consistent experimental conditions are maintained except for the number of training epochs, which is tailored to each scheme. Specifically, the DSC, STL, and MTL models are trained for $300$ epochs, while the CEP scheme utilizes $200$ epochs for channel estimation model training and $100$ epochs for precoding model training, both of which are observed to reach near convergence. We set the signal-to-noise ratio (SNR) as $\mathrm{SNR} = 10\log_{10}\!\left(\frac{P}{\sigma^2}\right)$ for all schemes. All other experimental settings not explicitly mentioned are kept uniform to ensure the validity of comparative evaluations. The strength of the underlying assumptions decreases across the considered schemes: ZF and WMMSE assume full, noiseless CSIT acquisition; CEP assumes a finite pilot length and noisy channels but perfect uplink feedback; DSC, STL, and MTL assume finite pilots and feedback, and noisy uplink and downlink channels.

\section{Numerical Results and Discussions}

In this section, we first evaluate the performance of the proposed multi-task pretraining framework. We then assess the scalability of the pretrained model to unseen tasks and its generalization across diverse scenarios. Finally, we evaluate the scaling behavior of the backbone model.

\subsection{Performance of Multi-task Pretraining}

\begin{table*}[ht]
  \centering
  \caption{\textsc{Comparison of different schemes in terms of parameter size, computational complexity (floating-point operations, millions), and validation performance (spectral efficiency, bps/Hz) across datasets is presented. Entries in \textbf{bold} indicate the best performance, while those with \underline{underline} denote the second-best performance.}}
  \includegraphics[width=0.95\linewidth]{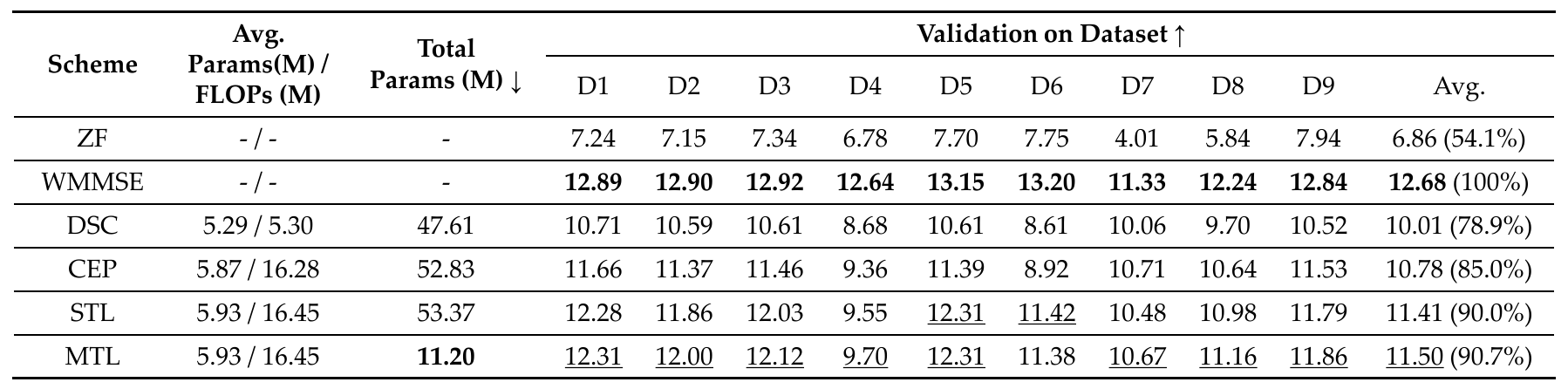}
  \label{tab:main_snr10}
\end{table*}

Table~\ref{tab:main_snr10} shows the validation spectral efficiency of the proposed scheme and baselines after pretraining at $\mathrm{SNR}=10$~dB. DL-based baselines (DSC, CEP, and STL) are trained separately for each Group~1 task, whereas ZF and WMMSE require no training. By contrast, the proposed MTL scheme is jointly optimized over all Group~1 tasks using a shared backbone. As shown in the table, the proposed method attains an average spectral efficiency of $11.50$~bps/Hz, exceeding DSC ($10.01$~bps/Hz) and CEP ($10.78$~bps/Hz) and remaining competitive with STL ($11.41$~bps/Hz). Meanwhile, the proposed MTL scheme substantially reduces the overall storage overhead, requiring only $11.20$~M parameters to serve all Group~1 tasks. This gain is primarily attributed to parameter sharing in the backbone, which enables models trained under heterogeneous configurations to reuse common precoder structures and thereby significantly reduces parameter redundancy.

\begin{figure}[h]
  \centering
  \includegraphics[width=0.9\linewidth]{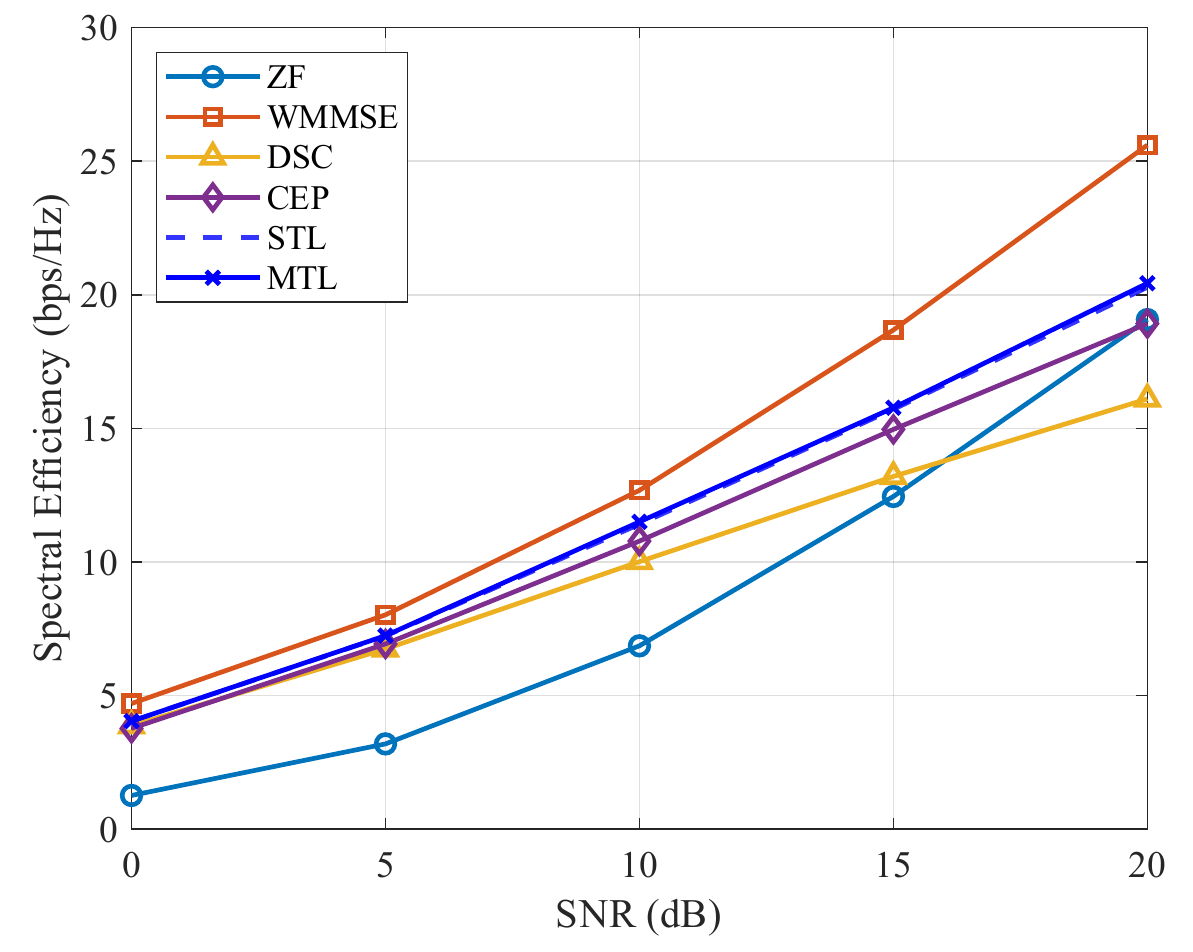}
  \caption{Comparison of the spectral efficiency of the proposed scheme and the baselines under different SNRs.}
  \label{fig:SE}
\end{figure}

In addition, the end-to-end schemes (the proposed method and STL) outperform the modular CEP pipeline ($11.50$ and $11.41$ vs.\ $10.78$~bps/Hz), despite CEP assumes perfect feedback and the end-to-end methods using finite feedback. This observation suggests that end-to-end optimization can alleviate error accumulation and objective mismatch across intermediate modules in modular design, thereby improving the effective spectral efficiency for FDD precoding. Figure~\ref{fig:SE} further compares the validation spectral efficiency under $\mathrm{SNR}\in\{0,5,10,15,20\}$~dB. Across the tested SNR range, the proposed MTL approach achieves the highest spectral efficiency among the evaluated baselines, indicating that the multi-task pretrained backbone remains effective under varying noise levels. Notably, all learning-based methods exhibit pronounced gains over ZF in the low-SNR regime, since they are trained to directly maximize the sum-rate objective and can learn noise-robust precoding behaviors. In contrast, ZF relies on channel inversion, which amplifies the impact of noise and leads to degraded performance at low SNR.

\subsection{Scalability to Unseen Tasks}

To evaluate the scalability of the model beyond the pretraining task set, we construct \textit{Group~2} (T10--T12) as a set of tasks that are unseen during pretraining. Compared with \textit{Group~1}, the differences mainly arise from task-defining system parameters (e.g., antenna array, pilot length and feedback overhead). This experiment is designed to address a focused question: whether the pretrained backbone can reuse precoding knowledge and adapt efficiently to new system configurations without large-scale retraining. Specifically, relative to \textit{Group~1}, \textit{Group~2} introduces (i) unseen antenna array settings (e.g., the $16\times 12$ UPA in T11), and (ii) unseen combinations of pilot/feedback ratios. These changes fundamentally redefine the task by altering both the signal dimensionality and the information bottleneck. The antenna array mainly affects the spatial degrees of freedom, while the pilot and feedback ratios determine the information budget available for channel acquisition and feedback.

\begin{figure}[h]
  \centering
  \includegraphics[width=0.9\linewidth]{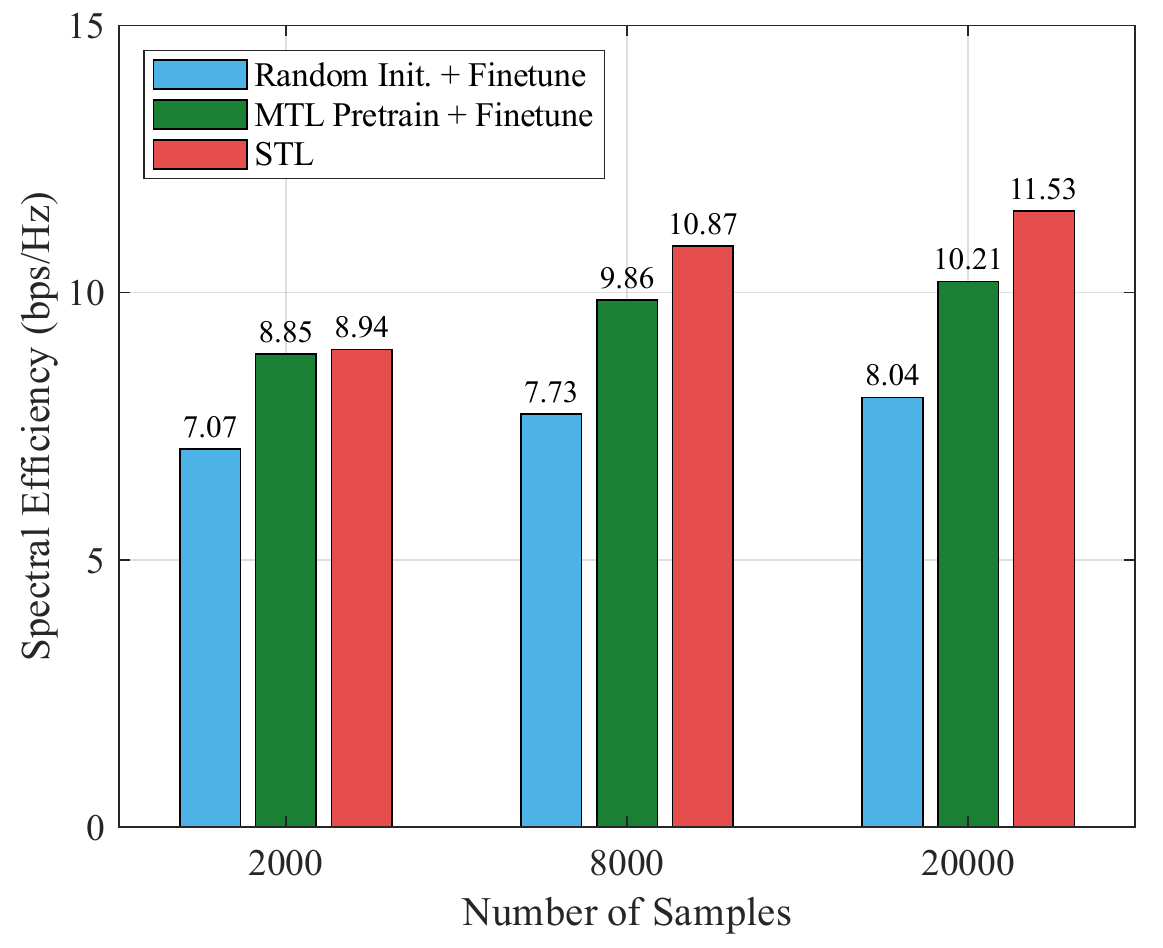}
  \caption{Fine-tuning performance on Group~2 tasks.}
  \label{fig:finetune}
\end{figure}

Figure~\ref{fig:finetune} compares three training strategies on \textit{Group~2}. (i) Random Init.\ + Fine-tune: the backbone is randomly initialized and frozen, and only a task-specific model is trained on \textit{Group~2}. (ii) MTL Pretrain + Fine-tune (proposed): the backbone is initialized by MTL pretraining on \textit{Group~1} and frozen, and only the task-specific model is trained on \textit{Group~2}. (iii) STL: the full model is trained on \textit{Group~2} with all parameters updated. The results show that the proposed scheme outperforms the randomly initialized one across all sample budgets (e.g., $8.85$ vs.\ $7.07$~bps/Hz at $2{,}000$ samples), indicating that the pretrained backbone captures reusable representations for precoding. In the few-shot regime ($2{,}000$ samples), the performance gap between the proposed scheme and STL is small ($8.85$ vs.\ $8.94$~bps/Hz), indicating that transfer provides a larger advantage with scarce target-task data.

\subsection{Domain Generalization}

\begin{table*}[h]
  \caption{\textsc{Domain Generalization performance.}}
  \includegraphics[width=0.95\linewidth]{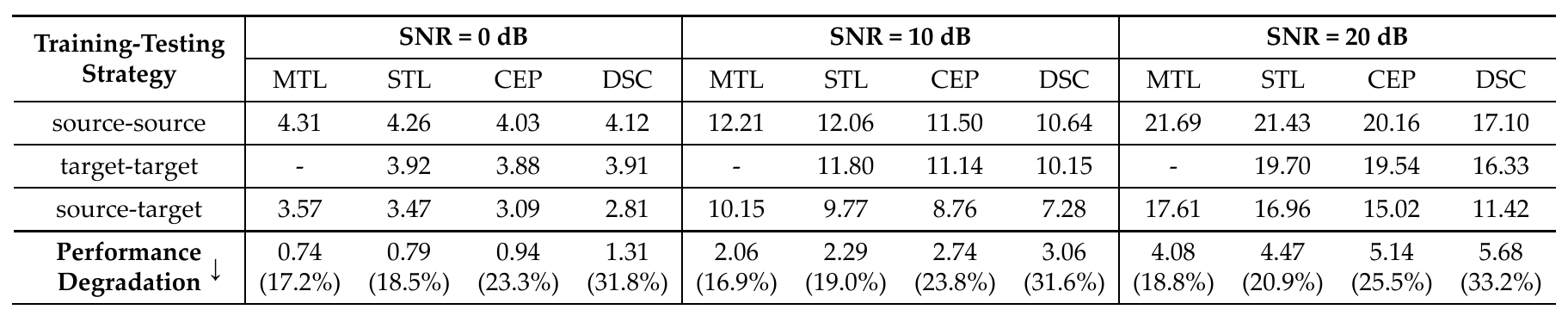}
  \label{tab:domain_gen}
\end{table*}

Table~\ref{tab:domain_gen} reports domain generalization under zero-shot transfer: models pretrained on \textit{Group~1} (source domain) are directly evaluated on \textit{Group~3} (target domain) without using any target-domain samples. In the table, ``source-source'' and ``source-target'' denote the average spectral efficiency on the source and target domains, respectively, where the average is taken over the target-domain task set (G1--G3) at a given SNR. The ``target-target'' results provide an in-domain training upper bound for reference. The generalization loss is measured by the absolute gap $D_{\rm sub}=L_{\rm source}-L_{\rm target}$ and the relative gap $D_{\rm div}=(L_{\rm source}-L_{\rm target})/L_{\rm source}$, where $L_{\rm source}$ and $L_{\rm target}$ denote the average performance across tasks on the source and target domains, respectively. Across all tested SNRs ($0/10/20$~dB), MTL achieves the highest target-domain spectral efficiency and the smallest generalization loss. At $\mathrm{SNR}=10$~dB, MTL reaches $10.15$~bps/Hz on the target domain with a degradation of $16.9\%$. STL, CEP, and DSC show larger degradations of $19.0\%$, $23.8\%$, and $31.6\%$, respectively. The same ordering holds at $\mathrm{SNR}=0$ and $20$~dB. While the absolute gap increases with SNR, the relative loss of MTL stays close across SNRs, indicating stable transfer.


This performance gain of MTL is attributed to the regularization and enhancement effects from the auxiliary tasks in \textit{Group~1}. Although the auxiliary tasks (T4--T9) are different from the target tasks, they are trained under the same multiuser precoding objective and therefore provide related system configurations and channel realizations to the shared backbone. By learning from related tasks, the backbone avoids focusing on patterns that only work for one specific source task and instead learns precoding structures that are useful in many tasks. As a result, the MTL scheme learns more stable precoding representations under changing channel statistics, which leads to easier transfer to the target domain and a smaller generalization loss than per-task training baselines.

\subsection{Model Scaling Effect}

\begin{figure}[h]
  \centering
  \includegraphics[width=0.9\linewidth]{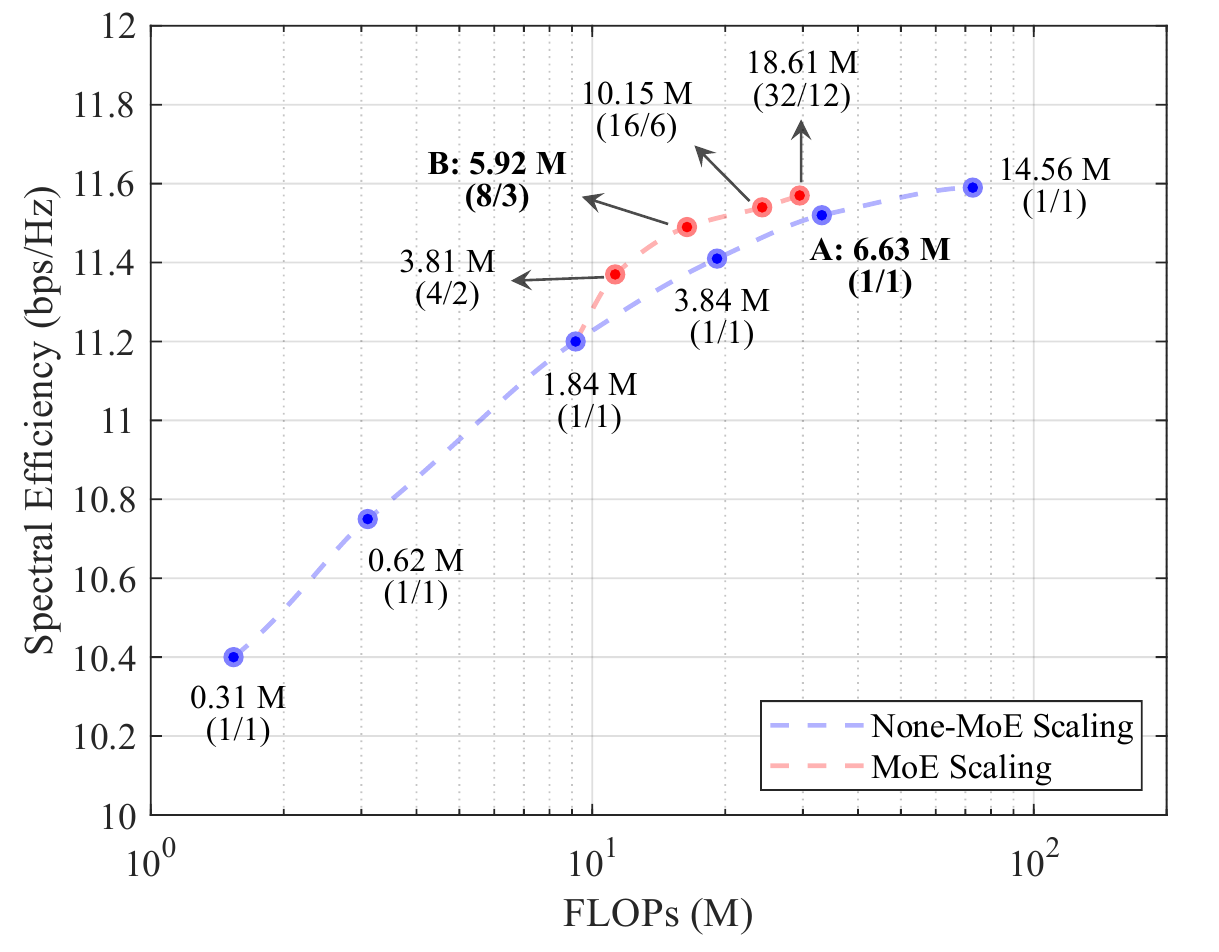}
  \caption{Model scaling performance.}
  \label{fig:model_scale}
\end{figure}

Figure~\ref{fig:model_scale} illustrates the trade-off between performance, storage, and floating-point operations (FLOPs) for different backbone architectures. The backbone based on a MoE-Transformer is scaled up via two distinct approaches: the non-MoE scaling method increases the dimensionality of the feed-forward and self-attention layers, whereas the MoE scaling method expands the number of experts. Each point in the figure corresponds to a specific model realization, annotated with its parameter count (in millions) and the total/activated number of experts. Due to the sparse activation property of MoE, under comparable FLOPs, models utilizing MoE scaling generally achieve superior performance compared to those relying solely on non-MoE scaling, thereby enabling a more favorable trade-off among performance, storage, and computational cost. For instance, Model~B requires approximately half the FLOPs and $0.71$ M fewer parameters than Model~A, while incurring only a $0.03$~bps/Hz degradation. This behavior is consistent with the design motivation of using a sparse MoE backbone: by performing conditional computation and dynamically routing each input to only a small subset of specialized experts, the MoE architecture increases the effective model capacity without proportionally increasing the computational cost, and mitigates task interference across heterogeneous configurations by separating shared and task-specific representations. In contrast, the non-MoE approach relies on a fully dense backbone, which lacks such routing-based inductive bias and is therefore less effective at capturing both task commonalities and distinctions.

\section{Conclusion}

In this paper, we proposed WiFo-E, a wireless foundation model for end-to-end precoding in FDD multi-user MIMO systems with heterogeneous configurations. By formulating precoder design under diverse configurations as a multi-task learning problem, WiFo-E leverages a shared  and task-specific networks to balance generalization and adaptability. Simulation results show that heterogeneous multi-task pretraining significantly outperforms independently trained models with lower storage cost and improved generalization to unseen tasks. Future work will extend this framework to other physical-layer tasks.


\bibliographystyle{IEEEtran}
\bibliography{references}

\end{document}